# Selective copper recovery from ammoniacal waste streams using a systematic biosorption process

Nina Ricci Nicomel[a,*], Lila Otero-Gonzalez[a,I], Adam Williamson[b], Yong Sik Ok[c], Pascal Van Der Voort[d], Tom Hennebel[b], Gijs Du Laing[a]

[a] Laboratory of Analytical Chemistry and Applied Ecochemistry, Department of Green Chemistry and Technology, Ghent University, Coupure Links 653, 9000 Ghent, Belgium

[b] Center for Microbial Ecology and Technology, Department of Biochemical and Microbial Technology, Ghent University, Coupure Links 653, 9000 Ghent, Belgium

[c] Korea Biochar Research Center, APRU Sustainable Waste Management Program and Division of Environmental Science and Ecological Engineering, Korea University, Seoul, Republic of Korea

[d] Center for Ordered Materials, Organometallics and Catalysis, Department of Chemistry, Ghent University, Krijgslaan 281 (S3), 9000 Ghent, Belgium

* Corresponding author:

E-mail: NinaRicci.Nicomel@UGent.be

Tel.: +3292646131

Fax: +3292646232

Present address:

[I] IDENER, Earle Ovington 24, 8-9, 41300 La Rinconada, Seville, Spain

**Highlights**

- A top-down approach optimized Cu biosorption from a real Cu-$NH_3$ leachate

- Pinecone (PC) has a high Cu $q_{max}$ of 1.1 mmol g$^{-1}$ despite 2 M $NH_3$ and 5 mM Zn present

- PC works over wide pH range (5-12) relevant for treating $(NH_4)_2CO_3$ rich effluents

- PC has a selectivity quotient ($K_{Cu/Zn}$) of 3.97 making it selective for Cu over Zn

- The presence of Zn can improve the $q_{max}$ of PC for Cu from $NH_3$-rich streams

**ABSTRACT**

Cu-NH$_3$ bearing effluents arise from electroplating and metal extraction industries, requiring innovative and sustainable Cu recovery technologies to reduce their adverse environmental impact. CO$_3^{2-}$ and Zn are often co-occurring, and thus, selective Cu recovery from these complex liquid streams is required for economic viability. This study assessed 23 sustainable biosorbents classified as tannin-rich, lignin-rich, chitosan/chitin, dead biomass, macroalgae or biochar for their Cu adsorption capacity and selectivity in a complex NH$_3$-bearing bioleachate. Under a preliminary screen with 12 mM Cu in 1 M ammoniacal solution, most biosorbents showed optimal Cu adsorption at pH 11, with pinecone remarkably showing high removal efficiencies (up to 68%) at all tested pH values. Further refinements on select biosorbents with pH, contact time, and presence of NH$_3$, Zn and CO$_3^{2-}$ showed again that pinecone has a high maximum adsorption capacity (1.07 mmol g$^{-1}$), worked over pH 5-12 and was Cu-selective with 3.97 selectivity quotient ($K_{Cu/Zn}$). Importantly, pinecone performance was maintained in a real Cu/NH$_3$/Zn/CO$_3^{2-}$ bioleachate, with 69.4% Cu removal efficiency. Unlike synthetic adsorbents, pinecones require no pre-treatment, which together with its abundance, selectivity, and efficiency without the need for prior NH$_3$ removal, makes it a competitive and sustainable Cu biosorbent for complex Cu-NH$_3$ bearing streams. Overall, this study demonstrated the potential of integrating bioleaching and biosorption as a clean Cu recovery technology utilizing only sustainable resources (i.e., bio-lixiviant and biosorbents). This presents a closed-loop approach to Cu extraction and recovery from wastes, thus effectively addressing elemental sustainability.

**Keywords:** copper, ammonia, adsorption, waste processing, selectivity

# 1. Introduction

Copper (Cu) and ammonia ($NH_3$) are two of the most used raw materials in many industrial processes, including printed circuit board (PCB) manufacturing and electroplating (Egenhofer et al., 2014; Al-Saydeh et al., 2017). Due to their extensive use, Cu and $NH_3$ are frequently found in the effluents of these processes at elevated levels. In PCB manufacturing, spent etching solutions can have Cu and $NH_3$ contents reaching up to 0.82 and 3.5 M, respectively, which present significant quantities of recoverable resources considering that the global PCB industry generates about $10^{12}$ L of waste etchant annually (Shah et al., 2018). Alternately, $NH_3$ has been recognized as an effective lixiviant in (bio)hydrometallurgical Cu extraction, resulting in Cu and $NH_3$ bearing liquid streams (i.e., leachates) with Cu concentrations exceeding 1.1 M (Xiao et al., 2013; Williamson et al., 2020). Considering the continuously growing global Cu demand (23.23 Mt in 2019) and the reduction in number of easily accessible primary Cu ores, Cu and $NH_3$-rich streams offer secondary resources to ensure sustainable Cu supply in the long run (Garside, 2020). However, these streams can be complex due to the co-occurrence of other components (e.g., Zn, $CO_3^{2-}$) and can have variable pH values mostly between pH 9 and 12, which can impact the Cu recovery process (Xiao et al., 2013; Chai et al., 2017; Williamson et al., 2020).

Cu recycling from aqueous secondary resources has been widely studied in the past to address resource availability concerns and to reduce the environmental burdens associated with primary Cu processing (Al-Saydeh et al., 2017; Ciacci et al., 2020). However, many Cu recycling studies have dealt with single metal solutions despite the multicomponent nature of real industrial effluents, which can potentially modify the type of interactions among the components and consequently affect the efficiency of the employed Cu recovery technology (Peng et al., 2011; Hu et al., 2017). The conventional Cu recovery technologies (i.e., hydroxide precipitation, ion exchange, membrane filtration, and electrochemical methods) are mostly inefficient in multicomponent solutions such as ammoniacal streams due to the formation of stable Cu-$NH_3$ complexes that are hard to break (Fu and Wang, 2011; Chai et al., 2017). Furthermore, their applications are usually hindered by high energy and chemical requirements, high capital costs, and the generation of voluminous by-products (Fu and Wang, 2011). Thus, alternative technologies are gaining attention for the sustainable and effective extraction, recovery and reuse of this economically important metal from secondary resources.

Adsorption has been considered as an alternative Cu recovery technology owing to its easy and low-cost operation, sustainability, and effectiveness even in complex solutions (Al-Saydeh et al., 2017). The systematic selection of adsorbent with high adsorption capacity and selectivity is essential to any adsorption process. Some adsorbents, including commercial silica gel, magnetic ferrite nanoparticles, and metal oxides (e.g., $TiO_2$ and $SnO_2$), have been investigated for Cu removal from ammoniacal solutions (Kar et al., 1975; Kar et al., 1976; Baba et al., 1984; Fuerstenau and Osseo-Asare, 1987). However, the unsustainable and costly materials used to synthesize these adsorbents are unjustified in environmental applications. Furthermore, most studies have used low $NH_3$ concentrations and solution pH ranges that are not representative of real Cu-$NH_3$ streams. The selection of these experimental conditions highlights the repetitive and limiting system employed in most adsorption studies, wherein a particular adsorbent is selected without an order and subsequently tested for different conditions with no target applications.

Alternatively, adsorbents from natural sources (i.e., biosorbents) can offer a sustainable and practical substitute for the synthetic adsorbents. To date, biosorbents have not been investigated for their capacity to remove Cu from ammoniacal streams, although they naturally possess surface functional groups, including carboxylic, amino, and phenolic-type surface groups, which have been reported to influence Cu adsorption (Gerente et al., 2007; He and Chen, 2014; Inyang et al., 2016). In some cases where ligands are present in the solution, the biosorbent efficiency and selectivity can even be enhanced if the ligand functional groups responsible for metal complexation are not involved in binding the ligand to the biosorbent surface (Elliott and Huang, 1979). Metal-biosorbent surface interaction is a complex and unpredictable process that is highly dependent on the environmental conditions. Thus, although many biosorbents have been previously studied for Cu adsorption, their efficiency and suitability for a particular stream cannot be directly compared given the differences in the experimental conditions used. A more holistic approach, such as a biosorption screening framework, reflecting the same range of conditions for all biosorbents would be more useful in generating information on adsorption capacity and selectivity.

While there has been an increasing number of biosorption-based Cu recovery studies using single-metal solutions, Cu recovery from complex solutions containing $NH_3$ and other ions is lacking. It is established that metal speciation changes in the presence of other ions and this largely influences metal adsorption. Recently, Williamson et al. (2020) has demonstrated that biogenic $NH_3$ can be a potent lixiviant for Cu and Zn extraction from heterogeneous automotive shredder residues (ASR), offering a greener lixiviant production route and Cu secondary resource. In order to close the loop and make such technology more feasible, the bioleaching process should be coupled with a technology that could efficiently and selectively recover Cu from the leachate despite the presence of $NH_3$ and other ions. In the present work, a set of 23 biosorbents was screened for the adsorption of Cu from a synthetic Cu-$NH_3$ leachate. An in-depth investigation of the adsorption of Cu-$NH_3$ complexes using five biosorbents (i.e., pinecone, *Fucus spiralis*, chitosan, sewage sludge biochar, and lignin-rich digested stillage) was subsequently performed by studying the influence of kinetics, solution pH, and presence of other ions (i.e., Zn, $CO_3^{2-}$) in the solution. Finally, a combination of optimal parameters and biosorbent was applied to the real Cu-$NH_3$ bearing leachate containing Cu, $NH_3$, $CO_3^{2-}$, and Zn (Williamson et al., 2020) to test our selection method and validate our findings.

## 2. Materials and Methods

*2.1. Biosorbents*

The biosorbents used in this study (Table 1) were selected based on positive reports in the literature identifying some of the most efficient Cu(II) biosorbents. These materials were classified according to the following biosorbent categories: bark/tannin-rich materials, lignin-rich materials, chitin/chitosan, dead biomass, macroalgae, and biochars (Bailey et al., 1999). The tannin-rich materials were obtained from natural sources including residues from woodworks, cacao production, and tea consumption. The lignin-rich agricultural bio-wastes wheat straw and hay were obtained commercially. The lignin-rich digested stillage was obtained from lignocellulosic ethanol production and subsequent anaerobic digestion (Ghysels et al., 2019). Chitosan with medium molecular weight was purchased from Sigma-Aldrich (St. Louis, MO, USA). Microalgal biomass was provided by GEMMA at Universitat Politècnica de Catalunya (Barcelona, Spain). The waste yeast was obtained from a local Belgian brewery. Macroalgal species were collected in Goes, The Netherlands with the help of the Phycology Research Group at Ghent University. Biochars were provided by the Korea Biochar Research Center. The biosorbents, except chitosan, were washed with deionized water several times and then dried in an oven at 60 °C for at least 48 h. The dried biosorbents were crushed and sieved to a particle size of <1 mm.

**Table 1**. List of the biosorbents used for the removal of Cu(II) from ammoniacal solution in the preliminary screening experiment.

| Biosorbent | Code | Description | Reference |
|---|---|---|---|
| *Bark/tannin-rich materials* | PC | Pinecone | Ofomaja et al. (2009) |
| | RCS | Roasted cacao shells | Meunier et al. (2003) |
| | SD | Sawdust | Rahman and Islam (2009) |
| | BTW | Black tea waste | Weng et al. (2014) |
| *Lignin-rich materials* | WS | Wheat straw | Han et al. (2010) |
| | HY | Hay[†] | Bailey et al. (1999) |
| | LRD | Lignin-rich digested stillage[†] | Bailey et al. (1999) |
| *Chitin/chitosan* | CT | Chitosan | Gerente et al. (2007) |
| | CS | Crab shells | Vijayaraghavan et al. (2006) |
| *Dead biomass* | WY | Waste yeast | Cojocaru et al. (2009) |
| | HC | Microalgae[†] | Bailey et al. (1999) |
| *Macroalgae* | SM | *Sargassum muticum*[†] | He and Chen (2014) |
| | FS | *Fucus spiralis* | Murphy et al. (2007) |
| | GG | *Gracilaria gracilis*[†] | He and Chen (2014) |
| | GT | *Grateloupia turuturu*[†] | He and Chen (2014) |
| | AS | *Agardhiella subulata*[†] | He and Chen (2014) |
| | UR | *Ulva rigida*[†] | He and Chen (2014) |
| | UC | *Ulva compressa* | Murphy et al. (2007) |
| | CF | *Codium fragile*[†] | He and Chen (2014) |
| *Biochar* | SSBC500 | Sewage sludge biochar[‡] | Otero et al. (2009) |
| | RHBC500 | Rice husk biochar[‡] | Pellera et al. (2012) |
| | PCBC200 | Pinecone biochar[†,*] | Inyang et al. (2016) |
| | PCBC500 | Pinecone biochar[†,‡] | Inyang et al. (2016) |

[†]References represent the closest literature match to the enlisted biosorbents; [‡]Pyrolyzed at 500 °C; [*]Pyrolyzed at 200 °C

*2.2. Screening of the biosorbents*

The biosorbents listed in Table 1 were screened for their Cu(II) removal efficiencies in the presence of $NH_3$ at different pH values (9, 10, 11 and 12). Four working solutions containing 12 mM Cu(II) as $Cu(NO_3)_2 \cdot 2.5H_2O$ (Chem-Lab NV, Belgium) and 1 M $NH_3$ (25 wt%, Chem-Lab NV, Belgium) were prepared by mixing known volumes of 200 mM Cu(II) stock solution and $NH_3$ solution, then diluted with deionized water to the correct volume. The pH of each solution was adjusted to a value from 9–12 with one pH unit increment using either concentrated $HNO_3$ or 10 M NaOH. Batch experiments were performed by adding 10 mL of the working solution to 100 mg of the biosorbent weighed in a 12-mL polypropylene tube (100 mL g$^{-1}$ liquid-to-solid (L/S) ratio). The samples were shaken using an orbital shaker at 115 rpm for 24 hours. After measuring the final pH using a Thermo Scientific Orion Star A211 pH meter, the suspensions were filtered using Whatman cellulose filter papers (11-µm pore size). The Cu concentration in the filtrates was determined by inductively coupled plasma optical emission spectroscopy (Varian Vista-MPX CCD Simultaneous ICP-OES). All experiments were conducted in duplicate and the data presented are mean values with standard deviation. Control samples were prepared without biosorbent to estimate the Cu(II) precipitation at different pH values. The Cu(II) removal efficiency and adsorption capacity of the biosorbents were calculated using Eq. A.1 and A.2 (Appendix A: Supplementary Data), respectively.

Based on the results of the screening, five biosorbents from different categories were selected for the succeeding experiments: pinecone (PC), *Fucus spiralis* (FS), chitosan (CT), sewage sludge biochar (SSBC500), and lignin-rich digested stillage (LRD). The effects of pH, contact time, $NH_3$ concentration, and presence of Zn(II) and $CO_3^{2-}$ on Cu(II) adsorption onto each of these biosorbents were investigated.

*2.3. Effect of pH on Cu(II) biosorption in the presence of $NH_3$*

Using PC, FS, CT, SSBC500, and LRD, the dependence of Cu(II) adsorption on the solution pH was investigated in batch experiments at two different $NH_3$ concentrations. Six solutions with pH of 3, 6, 7.5, 9, 11, and 12 were prepared containing 5 mM Cu(II) and either 1 or 2 M $NH_3$. The samples were prepared by adding 10 mL of the Cu-$NH_3$ solutions to 100 mg of biosorbent. After shaking the samples for 24 hours, all samples were filtered and analyzed for Cu concentration as described in Section 2.2.

*2.4. Cu(II) biosorption kinetics*

Kinetic studies were performed to investigate the adsorption of Cu(II) in the presence of $NH_3$ as a function of time. The biosorbents were put in contact with 10 mL of a solution (pH 11) containing 5 mM Cu(II) and 2 M $NH_3$ at an L/S ratio of 100 mL $g^{-1}$. The suspensions were shaken for different time intervals ranging from 15 minutes to 24 hours. After each time interval, the samples were filtered and analyzed for Cu concentration as described in Section 2.2.

*2.5. Biosorption isotherms*

The adsorption of Cu(II) on the selected biosorbents was assessed at different initial Cu concentrations to determine the maximum adsorption capacities ($q_{max}$) of the biosorbents at pH 11. The samples were handled in the same way as described in Section 2.2, except that the initial $NH_3$ concentration was 2 M, while the initial Cu(II) concentration varied from 5–100 mM. After filtering, the Cu concentrations of the filtrate were measured using ICP-OES. Langmuir (Eq. A.3) and Freundlich (Eq. A.4) isotherm models were used to fit the experimental data to estimate the maximum adsorption capacities of the biosorbents. The adsorption isotherms and the corresponding parameters were determined using SigmaPlot v13.0.

*2.6. Selectivity between Cu(II) and Zn(II)*

Batch experiments were conducted to investigate competition effects and selectivity between Cu(II) and Zn(II). A solution containing 50 mM Cu(II), 50 mM Zn(II) as $Zn(NO_3)_2·6H_2O$ (Merck, Germany), and 2 M $NH_3$ was prepared at pH 11. The initial Cu(II) concentration used was based on the concentration where the biosorbents exhibited saturation, while Zn(II) was set to have an equimolar concentration. When biosorbents are saturated, the competition for adsorption sites is maximized, leading to correct assessment of the selectivity between Cu(II) and Zn(II). The samples were prepared and analyzed for Cu and Zn concentration by ICP-OES as described in Section 2.2. Selectivity quotients ($K_{Cu/Zn}$), given by Eq. A.5 (Kar et al., 1976), were calculated to compare the ability of the biosorbents to selectively adsorb Cu(II) over Zn(II) from a system of 50 mM Cu(II), 50 mM Zn(II), and 2 M $NH_3$.

The $q_{max}$ of the five biosorbents for Cu(II) were re-evaluated in the presence of Zn(II) to assess how Zn(II) will impact the $q_{max}$ values obtained in a single-metal solution. This experiment was set up as described in Section 2.5, but using a multimetal solution containing both Zn(II) and Cu(II). The initial pH used was pH 11 and the amount of Zn(II) added to the solution was 5 mM.

*2.7. Effect of carbonate on Cu(II) biosorption*

The effect of $CO_3^{2-}$ on the adsorption of Cu(II) was investigated in batch experiments. Two solutions with initial pH of 9 and 11 were prepared, each containing 5 mM Cu(II), 2 M $NH_3$, and 1 M $CO_3^{2-}$ added as $(NH_4)_2CO_3$ ($\geq$ 30% $NH_3$ basis, Sigma-Aldrich, USA). The samples were handled and analyzed for Cu concentration as described in Section 2.2.

*2.8. Cu(II) biosorption from real leachates*

PC, FS, CT, SSBC500, and LRD were tested for Cu(II) adsorption from a real leachate (pH 11.1) derived from the extraction of ASR, which contained 6 mM Cu(II), 1.1 mM Zn(II), 1 M $NH_3$, and 0.5 M $CO_3^{2-}$ (Williamson et al., 2020). A simplified synthetic leachate with approximately the same composition was prepared as a reference. The samples were prepared and analyzed as described in Section 2.2.

3. Results and Discussion

*3.1. Screening of the biosorbents*

An initial screen of 23 biosorbents was performed between pH 9-12 to determine the optimal biosorbent and pH for the biosorption of Cu(II) from a solution of 12 mM Cu(II) and 1 M $NH_3$ (Fig. 1). No systematic biosorption trend within the same category was observed. Most biosorbents showed poor (< 20%) Cu(II) adsorption at pH 9 and 10, with Cu speciation (Fig. A.1) modeled solely as $Cu(NH_3)_4^{2+}$. In general, Cu(II) removal efficiencies were higher at pH 11 (increase < 15%), where both $Cu(NH_3)_4^{2+}$ and $Cu(NH_3)_3OH^+$ are present in equal fractions. Enhanced adsorption at pH 11 could be attributed to the presence of the hydroxide ligand, which can increase the probability of adsorption through interfacial hydrogen bonding with the surface functional groups of the biosorbents (Fuerstenau and Osseo-Asare, 1987; Crawford et al., 1997). Furthermore, the enhanced adsorption at higher pH values can be attributed to increased coulombic interaction (Fuerstenau and Osseo-Asare, 1987), with negatively charged surfaces attracting $Cu(NH_3)_4^{2+}$ and $Cu(NH_3)_3OH^+$. At pH 12, a removal of 25% was observed in the control, likely due to the formation of $Cu(OH)_2$ precipitates. After correcting for this, the Cu(II) adsorption was generally less at pH 12 than at pH 11, with more than half of the biosorbents showing reduced removal efficiencies (< 25%).

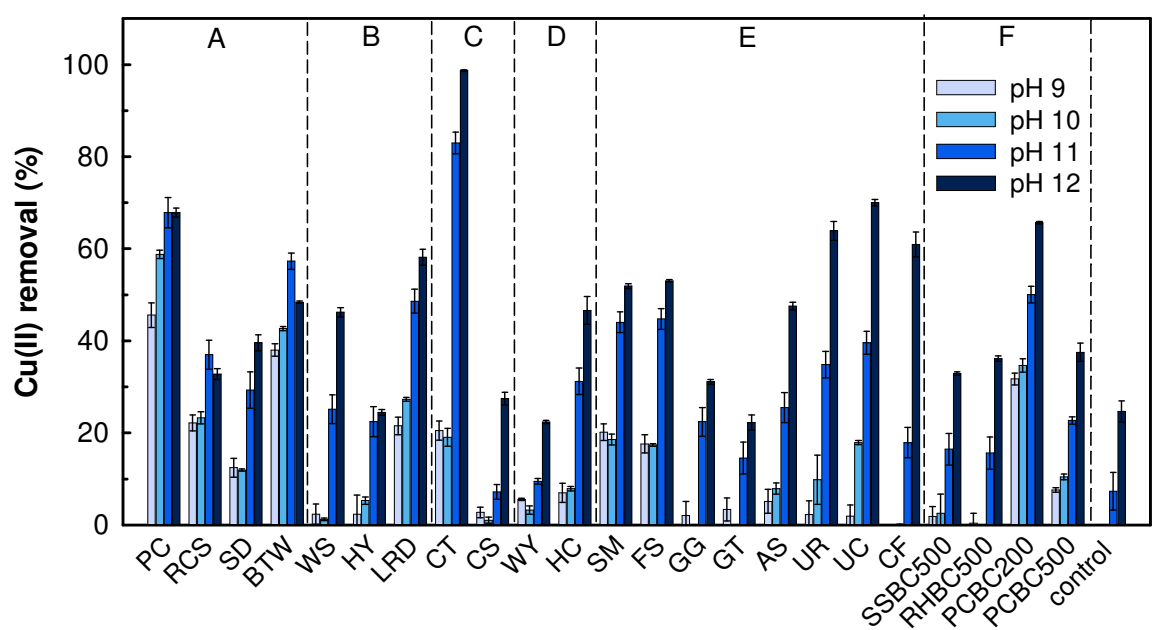

**Fig. 1.** Cu(II) removal efficiencies of different biosorbents in ammoniacal solutions with different pH values. Experimental conditions: 12 mM Cu(II), 1 M $NH_3$, 100 mL g$^{-1}$ L/S ratio, room temperature (RT, 20 ± 2 °C), 24 h contact time. Note: The biosorbents were grouped according to the categories (A) tannin-rich materials, (B) lignin-rich materials, (C) chitin/chitosan, (D) dead biomass, (E) macroalgae, and (F) biochar.

This preliminary experiment provided a first approximation of the Cu(II) adsorption behavior of the biosorbents in the presence of $NH_3$. Based on the results, five biosorbents were selected for further study, namely PC, FS, CT, SSBC500, and LRD. The tannin-rich PC showed relatively good removal efficiencies regardless of the pH and was the most efficient biosorbent at pH 9 (46%) and 10 (59%). CT had the highest Cu(II) removal at pH 11 (83%), while the alginate-rich FS and lignin-rich LRD both have acceptable removal efficiencies at pH 11. While exhibiting poor Cu affinities, SSBC500 was selected to identify which surface properties are unfavorable for Cu(II) adsorption. PC, FS, CT, SSBC500, and LRD were characterized by performing point of zero charge ($pH_{PZC}$) determination, pore structure characterization, acidic and basic surface properties determination (Table A.1), and FTIR spectroscopy (Table A.2). These biosorbents and any similar ones have not been reported to remove Cu-$NH_3$ complexes before. Most studies concerning Cu(II) adsorption from ammoniacal solutions used commercial silica gel and synthetic or natural oxides (Kar et al., 1975; Kar et al., 1976; Baba et al., 1984; Fuerstenau and Osseo-Asare, 1987).

*3.2. Effect of pH and $NH_3$ concentration on Cu(II) removal*

Metal speciation varies over a wide pH range and this has an important role in optimizing metal adsorption processes. Thus, Cu(II) adsorption (5 mM) by PC, FS, CT, SSBC500, and LRD was assessed from pH 3 to 12 in the presence of 1 M or 2 M $NH_3$ to determine the influence of Cu(II) speciation on the optimum pH for Cu(II) adsorption (Fig. 2). Except that of SSBC500, Cu(II) adsorption increased by 4–72% as the pH increased from 3 to 6. A drop in Cu(II) adsorption of up to 95% was observed between pH 6 and 9, followed by another increase from pH 9 to 12. Furthermore, as the $NH_3$ concentration increased from 1 to 2 M, only a small decrease in the Cu(II) removal efficiencies was observed, except for CT which showed a decrease of 55% at pH 11. Metal adsorption capacities decrease in the presence of excess $NH_3$, which compete with metal ions for the biosorbent surface sites (Kar et al., 1976).

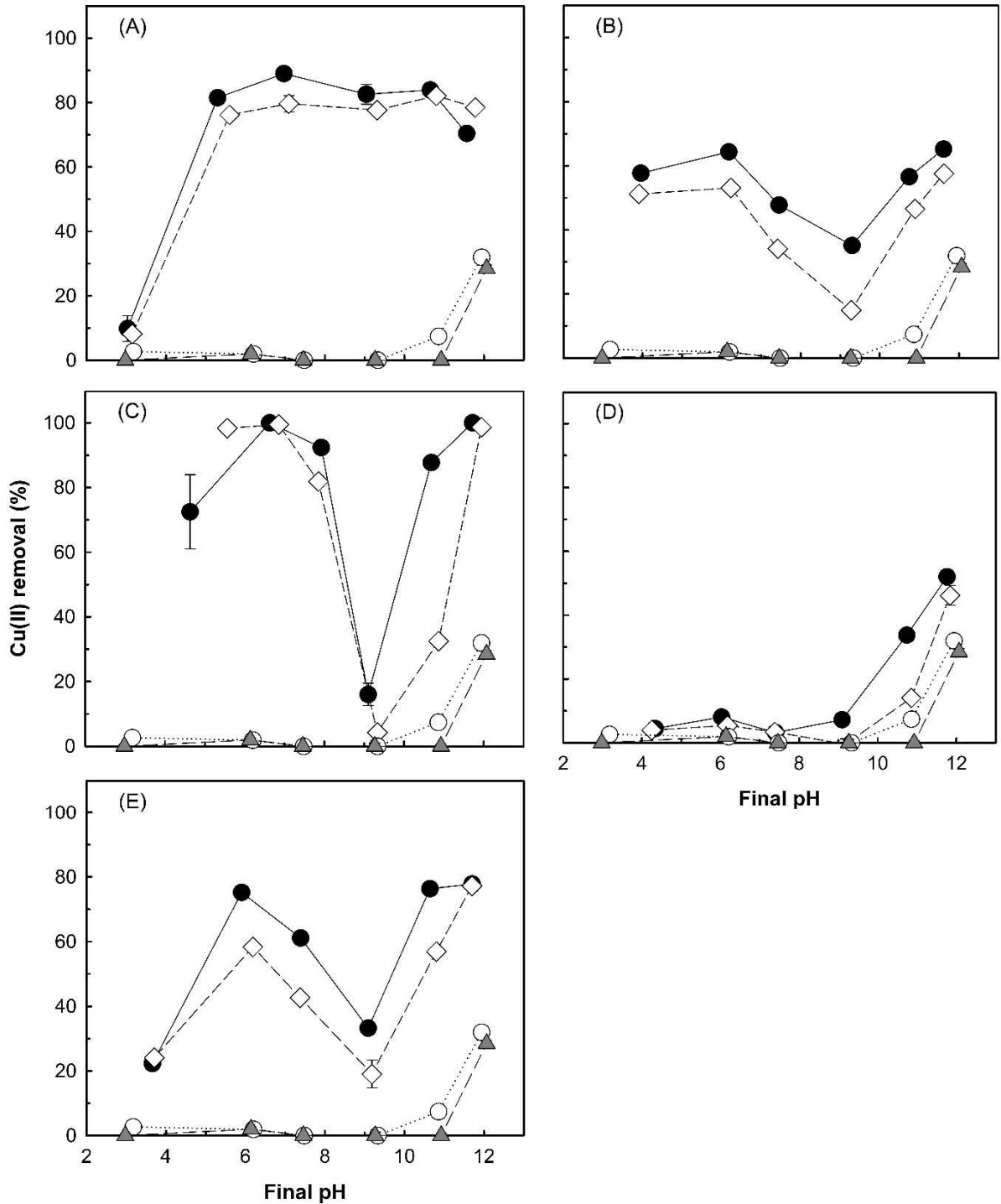

**Fig. 2.** Effect of pH on the Cu(II) removal efficiencies of (A) PC, (B) FS, (C) CT, (D) SSBC500, and (E) LRD in the presence of (●) 1 M and (◊) 2 M $NH_3$. Cu(II) removal by the control samples at (○) 1 M and (▲) 2 M $NH_3$ were also indicated. Experimental conditions: 5 mM Cu(II), 100 mL $g^{-1}$ L/S ratio, RT (20 ± 2 °C), 24 h contact time. Note: The straight lines were added for visual clarity only.

At pH 3, low Cu(II) removal efficiencies (< 25%) were observed for PC, SSBC500, and LRD (Figs. 2A, 2D and 2E), which could be due to electrostatic repulsion between the predominant $Cu^{2+}$ (Fig. A.2) and the positively charged biosorbents surface as suggested by their $pH_{PZC}$ (Table A.1). FS and CT achieved relatively high removal efficiencies (> 50%) at pH 3 (Figs. 2B and 2C), possibly because of the deprotonation of some functional groups specific for these biosorbents. FS contains sulfonic acid surface groups as shown by the FTIR analysis (Table A.2). These functional groups dissociate to form negatively charged sulfonate groups when pH is higher than its $pK_a$ of 0.5 (van Loon et al., 1993).

At pH 6, CT showed the highest Cu(II) removal of 100% (Fig. 2C), but this could have been influenced by its relatively higher final pH of 6.7 compared to those of the other biosorbents. FS, CT, and LRD achieved their highest Cu(II) removal efficiencies at around pH 6, where the sum of the fractions of the major species $Cu(NH_3)^{2+}$, $Cu(NH_3)_2^{2+}$, and $Cu(NH_3)_3^{2+}$ reaches its maximum (Fig. A.2). Baba et al. (1984) observed a similar trend as Cu(II) adsorption from ammoniacal solutions on silica gel increased until pH 6, and then decreased thereafter.

At pH 7.5, the removal efficiencies of FS, CT, and LRD started to decrease by 7.7-18.9%, while that of PC remained almost constant. At pH 9, most biosorbents reached their minimum Cu(II) removal efficiencies (< 35%), except PC which achieved 82.6% (1 M $NH_3$) and 77.7% (2 M $NH_3$) (Fig. 2A) indicating further that Cu-$NH_3$ adsorption on PC is less pH-dependent compared to the other biosorbents. Fuerstenau and Osseo-Asare (1987) also reported minimum Cu(II) adsorption on hematite at pH 9.25 after reaching a maximum at pH 7. $Cu(NH_3)_4^{2+}$ is the predominant Cu(II) species at pH 9. The Cu(II) ion has six coordination sites: two axial sites and four sites in the square plane for the attachment of up to four $NH_3$ ligands. In the complexes $Cu(NH_3)^{2+}$, $Cu(NH_3)_2^{2+}$, and $Cu(NH_3)_3^{2+}$, two axial sites and at least one site in the square plane are available, whereas in $Cu(NH_3)_4^{2+}$, only the two axial sites are available (Sharma et al., 2011). This suggests that Cu-$NH_3$ complexes with at least one free coordination site in the square plane are more favorably adsorbed. This is in accordance with the study of Tominaga et al. (1975), in which electron spin resonance spectroscopy revealed that the Cu species adsorbed on silica gel from an ammoniacal solution were neither $Cu^{2+}$ nor $Cu(NH_3)_4^{2+}$, but $Cu(NH_3)_n^{2+}$ (n = 1–3). While minimum Cu(II) removal at around pH 9 has been previously reported, the consistent Cu(II) removal potential of PC between pH 5 and 12 has not been reported to date and clearly holds an advantage over other biosorbents for its application towards a broader spectrum of ammoniacal Cu-bearing wastes.

In general, the highest Cu(II) adsorption in the basic pH range 9-12 was achieved at pH 11 considering the removal observed in the control at pH 12 (i.e., 33% at 1 M $NH_3$ and 29% at 2 M $NH_3$). This could be due to the hydroxide ligand(s) present in the predominant hydroxyammine Cu complexes (Fig. A.2). The extent of drop and rebound is different for each biosorbent suggesting that it is dependent on the biosorbent characteristics. Crawford et al. (1997) observed this in a similar system and noted that the hydroxide ligand is essential to metal ion adsorption as it can participate in chemical interaction (e.g., hydrogen bonding) with the adsorbent surface.

*3.3. Kinetics of Cu(II) biosorption in the presence of $NH_3$*

The results in Figs. 1 and 2 showed that within the typical pH range of ammoniacal waste streams (i.e., pH 9-12), the optimum Cu(II) removal efficiency without the influence of precipitation can be achieved at pH 11. Hence, kinetic experiments were performed at pH 11 to determine the rate of Cu(II) adsorption. Equilibrium was attained within 30 min in the case of FS, CT, and LRD, 1 h in PC, and 8 h in SSBC500 (Fig. 3). Rapid initial adsorption was observed for FS, CT, and LRD (48-75%), followed by a slow increase over 16 hours. PC showed the highest Cu(II) removal efficiency (75%) after a 15-min contact time. Like in the case of SSBC500, slow adsorption of some metal-ammine complexes including Cu-$NH_3$ was previously reported to be caused by chemical changes in the adsorbed layer following the initial adsorption of the complex ion, rather than to a low adsorption rate (Smith, 1939).

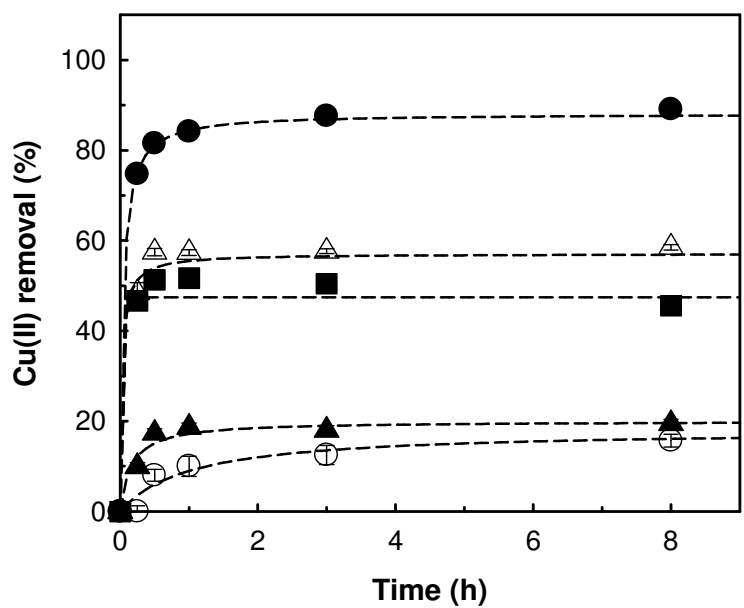

**Fig. 3.** Kinetics of Cu(II) biosorption on PC (●), FS (■), CT (▲), SSBC500 (○), and LRD (△) in ammoniacal solution. Experimental conditions: 5 mM Cu(II), 2 M $NH_3$, 100 mL $g^{-1}$ L/S ratio, pH 11, RT (20 ± 2 °C).

*3.4. Biosorption isotherms*

The adsorption equilibrium provides information about the affinity of the adsorbent for the adsorbate, and its distribution in the liquid and solid phase (Lodeiro et al., 2008). The equilibrium data of the biosorbents were fitted to the Langmuir and Freundlich isotherm models (Fig. A.3). Table 2 shows the constants and correlation coefficients ($R^2$) from the fittings of these models. PC obtained the highest Langmuir adsorption equilibrium constant (*b*) of 1.101 L $mmol^{-1}$, indicating that it has more affinity for Cu compared to the other biosorbents.

Table 2. Constants and correlation coefficients of Langmuir and Freundlich models for Cu(II) biosorption in the presence of $NH_3$ using PC, FS, CT, SSBC500, and LRD. Experimental conditions: 2 M $NH_3$, 100 mL $g^{-1}$ L/S ratio, pH 11, RT (20 ± 2 °C), 24 h contact time.

| Biosorbent | Without Zn(II) | | | | | | With 5 mM Zn(II) | | | | | |
|---|---|---|---|---|---|---|---|---|---|---|---|---|
| | Langmuir model | | | Freundlich model | | | Langmuir model | | | Freundlich model | | |
| | $R^2$ | $b$ (L $mmol^{-1}$) | $q_{max}$ (mmol $g^{-1}$) | $R^2$ | $K_f$ ((mmol $g^{-1}$)(L $mmol^{-1}$)$^{1/n}$) | $n$ | $R^2$ | $b$ (L $mmol^{-1}$) | $q_{max}$ (mmol $g^{-1}$) | $R^2$ | $K_f$ ((mmol $g^{-1}$)(L $mmol^{-1}$)$^{1/n}$) | $n$ |
| PC | 0.988 | 1.101 | 0.94 | 0.765 | 0.528 | 7.30 | 0.966 | 0.548 | 1.07 | 0.793 | 0.239 | 3.47 |
| FS | 0.937 | 0.142 | 0.91 | 0.793 | 0.239 | 3.47 | 0.987 | 0.040 | 1.14 | 0.945 | 0.118 | 2.16 |
| CT | 0.967 | 0.059 | 2.03 | 0.885 | 0.235 | 2.19 | 0.963 | 0.006 | 2.34 | 0.951 | 0.019 | 1.20 |
| SSBC500 | 0.944 | 0.089 | 0.30 | 0.820 | 0.063 | 3.04 | 0.968 | 0.066 | 0.27 | 0.886 | 0.046 | 2.69 |
| LRD | 0.989 | 0.186 | 0.87 | 0.885 | 0.266 | 3.81 | 0.996 | 0.077 | 0.84 | 0.939 | 0.157 | 2.79 |

The maximum Cu(II) adsorption capacities ($q_{max}$) of the biosorbents at a final pH of 11 were determined from the Langmuir model (Table 2). The order of $q_{max}$ (mmol g$^{-1}$) follows: CT (2.03) > PC (0.94) > FS (0.91) > LRD (0.87) > SSBC500 (0.30). These $q_{max}$ values are of the same order of magnitude as those of some adsorbents reported in the literature. However, the differences in the adsorption systems used should be compared to evaluate the reported $q_{max}$ objectively. For instance, silica gel has a Cu $q_{max}$ of 2 mmol g$^{-1}$ at pH 6 in the presence of 6 M NH$_3$ (Smith and Jacobson, 1956). Graphite oxide (GO) has a Cu $q_{max}$ of 22 mmol g$^{-1}$ in a system with 0.6 M NH$_3$ and pH 11.2 (Kovtjukhova and Karpenko, 1992). This $q_{max}$ is higher by at least 14 times than those of the biosorbents used in this study, however, GO was not used in aqueous systems with much higher NH$_3$ concentrations. Furthermore, the particle sizes of GO can be as small as 0.1 µm (Taha-Tijerina et al., 2015), which are difficult to separate from the solution after adsorption. Lastly, the synthesis of GO involves oxidation with KMnO$_4$ in concentrated H$_2$SO$_4$, making this material expensive and unsustainable.

*3.5. Competition and selectivity between Cu(II) and Zn(II)*

Single metal systems rarely occur in real waste streams, making it important to evaluate the biosorbents in a competitive metal adsorption process. Since Cu(II) and Zn(II) coexist in the real leachate under study and other process and waste streams (e.g., electroplating wastewater), their adsorption characteristics in a binary metal system at equimolar concentrations (i.e., 50 mM) were investigated (Fig. 4). The Cu(II) adsorption capacity ($q_{Cu}$) of PC slightly increased by 0.07 mmol g$^{-1}$ when 50 mM Zn(II) was present, indicating different sorption sites for Cu and Zn. For the other biosorbents, FS, CT, SSBC500 and LRD, a decrease in $q_{Cu}$ was observed, but to variable degrees (0.03–0.89 mmol g$^{-1}$, Fig. 4). In the case of FS, competition was almost equivalent, with a $q_{Cu}$ of 0.36 mmol g$^{-1}$ and $q_{Zn}$ of 0.30 mmol g$^{-1}$. For CT, $q_{Cu}$ of 1.47 mmol g$^{-1}$ decreased to 0.58 mmol g$^{-1}$, although $q_{Zn}$ was only 0.13 mmol g$^{-1}$, indicating that one mole of Zn interacts with more functional groups than one mole of Cu. Multidentism could be the binding mechanism of Zn(II) to CT, which implies that more than one acidic group on the surface of CT was involved in binding to a single Zn ion (Lodeiro et al., 2008).

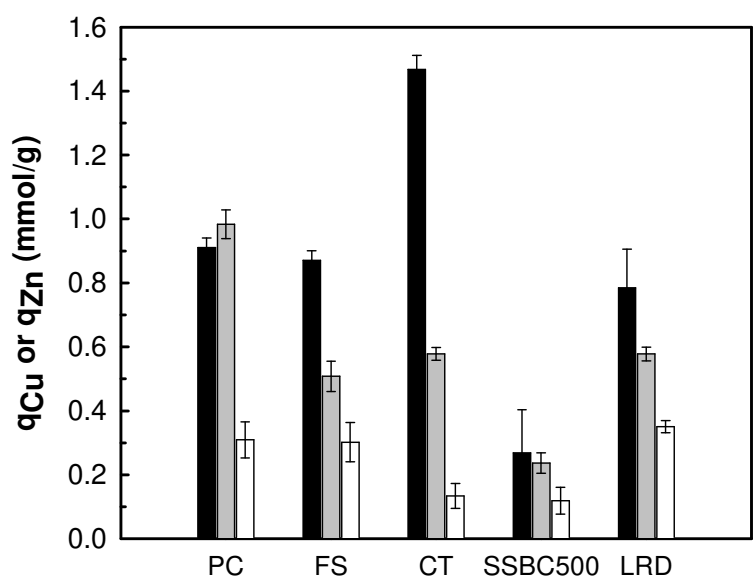

**Fig. 4.** Cu(II) (■) and Zn(II) (□) adsorption capacities of PC, FS, CT, SSBC500, and LRD in binary metal solutions containing $NH_3$. The Cu(II) adsorption capacities of the biosorbents in single metal ammoniacal solutions (■) are also shown in this figure for reference. Experimental conditions: 50 mM Cu(II), 0 mM Zn(II) (single metal solution) or 50 mM Zn(II) (binary metal solution), 2 M $NH_3$, 100 mL g$^{-1}$ L/S ratio, pH 11, RT (20 ± 2 °C), 24 h contact time.

Despite the decrease in $q_{Cu}$, all biosorbents exhibited higher affinity for Cu(II)-$NH_3$ complexes than Zn(II)-$NH_3$ complexes (Fig. 4). The selectivity quotients ($K_{Cu/Zn}$) support the higher affinity for Cu-$NH_3$ considering that all values are greater than 1—PC (3.97), FS (1.85), CT (4.98), SSBC500 (2.14), and LRD (1.84). This finding contrasts with the results of Kar et al. (1975) where hydrous β-$SnO_2$ was more selective for Zn(II)-$NH_3$ over Cu-$NH_3$ at pH 8.5. However, the selectivity quotient $K_{Zn/Cu}$ decreased from 2.41 to 1.08 as the equimolar concentrations of Zn(II) and Cu(II) increased from 0.01 to 0.15 M.

The results of the present study and those of Kar et al. (1975) suggest that the speciation of the Cu(II)-Zn(II)-$NH_3$ system (Fig. A.4) could influence the biosorbents' selectivity. At pH 11, the predominant Cu(II) species were $Cu(NH_3)_4^{2+}$ and $Cu(NH_3)_3OH^+$ in equal fractions, whereas for Zn(II), it was solely $Zn(NH_3)_4^{2+}$. Mixed hydroxyammine metal complexes (e.g., $Cu(NH_3)_3OH^+$) have higher possibility of being adsorbed because of the presence of at least one hydroxide ligand. Additional experiments showed that Cu(II) was still preferentially adsorbed over Zn(II) at pH 9 (Fig. A.5), despite both Cu(II) and Zn(II) tetraammine complexes form at this pH. Thus, aside from metal speciation, the complex ion orientation on the biosorbent surface could have affected the adsorption. The number of hydrogens that are spaced properly to form hydrogen bonds with the biosorbent surface oxygen atoms is important in complex ion adsorption (Smith and Jacobson, 1956). It is possible that the Cu(II)-ammine complexes attached to the biosorbent surface on their edge, resulting in more available adsorption sites, and thus, higher amounts of Cu(II)-ammine complex adsorbed.

The $q_{max}$ of the biosorbents for Cu(II) were also re-evaulated in the presence of Zn(II). Similar to the previous $q_{max}$ determination, a better fit of the equilibrium data of all biosorbents was obtained using the Langmuir model (Table 2). With 5 mM Zn(II) present, the order of $q_{max}$ (mmol g$^{-1}$) changed to: CT (2.34) > FS (1.14) > PC (1.07) > LRD (0.84) > SSBC500 (0.27). The $q_{max}$ of PC, FS, and CT for Cu(II) increased by 0.13, 0.23, and 0.31 mmol g$^{-1}$, respectively. The increase in the $q_{max}$ of PC was expected considering that the $q_{Cu}$ of PC increased despite the presence of 50 mM Zn(II) in the system (Fig. 4).

*3.6. Effect of carbonate on Cu(II) biosorption from ammoniacal solution*

The leachate from ASR extraction contains $CO_3^{2-}$, and thus, its effect on Cu(II) biosorption was investigated in systems containing 5 mM Cu(II), 2 M $NH_3$, and 1 M $CO_3^{2-}$ at pH 9 and 11. Generally, Cu(II) removal by all biosorbents decreased in the presence of $CO_3^{2-}$ (Fig. 5) with the decrease being more pronounced (~30%) at pH 11 for FS, CT and LRD. Cu(II) removal by PC also decreased, but it still achieved 68.6% removal. At pH 9, the decrease in Cu(II) removal efficiencies of most biosorbents was minimal, except that of PC, which decreased by 20.7%. Despite the presence of 1 M $CO_3^{2-}$, the predominant Cu(II) species remain as $Cu(NH_3)_4^{2+}$ (pH

9) and a mix of $Cu(NH_3)_4^{2+}$ and $Cu(NH_3)_3OH^+$ (pH 11) (Fig. A.6). Thus, the decrease in adsorption cannot be attributed to alteration of Cu(II) speciation, but rather likely due to the high amount of $CO_3^{2-}$ in the system. There have been reports regarding possible enhancement of metal and nonmetal adsorption in the presence of low $CO_3^{2-}$ concentrations through the formation of extra reactive sites coexisting with the adsorbed $CO_3^{2-}$ (Villalobos et al., 2001; Wijnja and Schulthess, 2002). However, at much higher $CO_3^{2-}$ concentrations, steric interactions may occur, leading to decreased metal adsorption (LaFlamme and Murray, 1987; Wijnja and Schulthess, 2002). The difference between the adsorption performances of the biosorbents could thus be due to $CO_3^{2-}$ speciation (Fig. A.6).

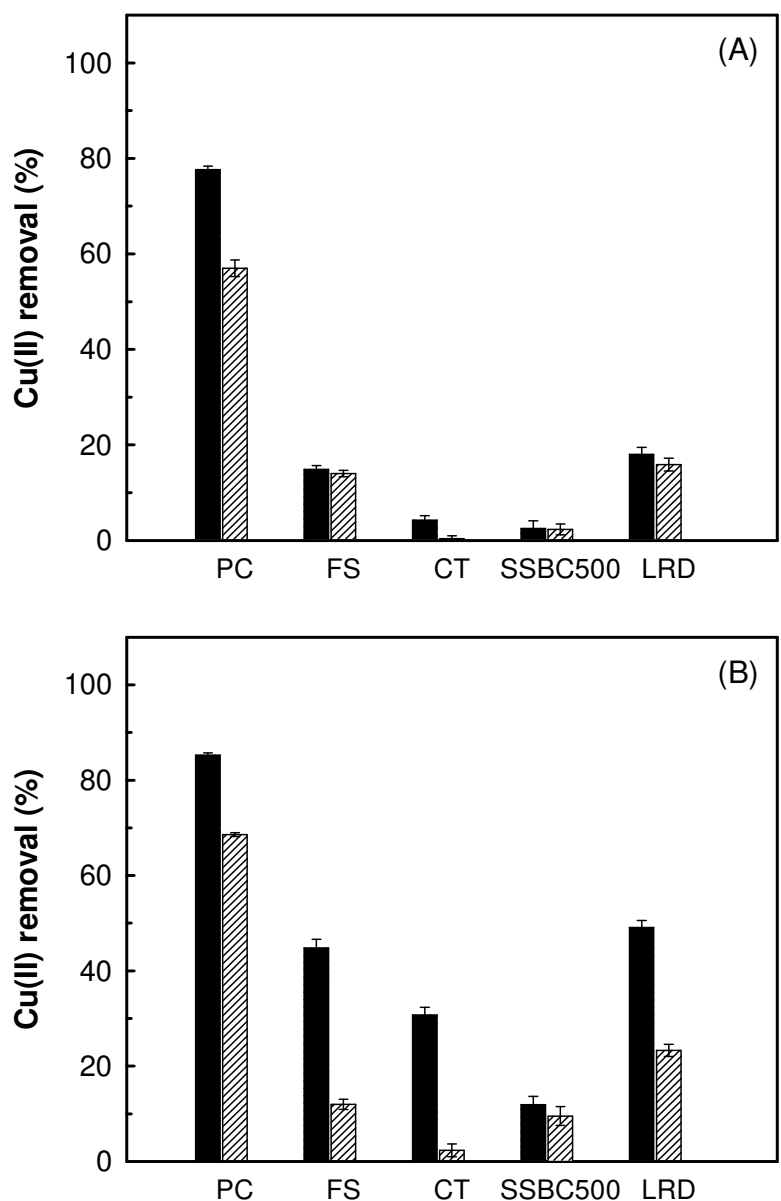

**Fig. 5.** Cu(II) removal of PC, FS, CT, SSBC500, and LRD from ammoniacal solution in the absence (■) and presence (▨) of $CO_3^{2-}$ at (A) pH 9 and (B) pH 11. Experimental conditions: 5 mM Cu(II), 2 M $NH_3$, 1 M $CO_3^{2-}$, 100 mL $g^{-1}$ L/S ratio, RT (20 ± 2 °C), 24 h contact time.

*3.7. Cu(II) biosorption from real Cu-NH₃ leachates*

The biosorbents were tested for their performance with a real leachate (pH 11.1) containing 6 mM Cu(II), 1.1 mM Zn(II), 1 M $NH_3$, and 0.5 M $CO_3^{2-}$ (Williamson et al., 2020). A simplified synthetic leachate with approximately the same composition was prepared as a reference and to highlight any other factors that impact biosorption. The Cu(II) removal efficiencies of the biosorbents used in the real leachate broadly matched the synthetic leachate (Fig. 6). The slight decrease (< 5%) in Cu(II) removal could be partially accounted to slightly higher initial Cu(II) concentration of the real leachate, or other uncharacterized organic/inorganic species that occurred during the leaching process. The results suggest that the performances of the biosorbents are comparable when used in both types of leachate, hence, validating that the results obtained in the previous sections are representative for real leachates. PC achieved the highest Cu(II) removal efficiency of 69.4% despite the presence of ligands and other ions in the real leachate. All other biosorbents achieved <30% Cu(II) removal efficiencies.

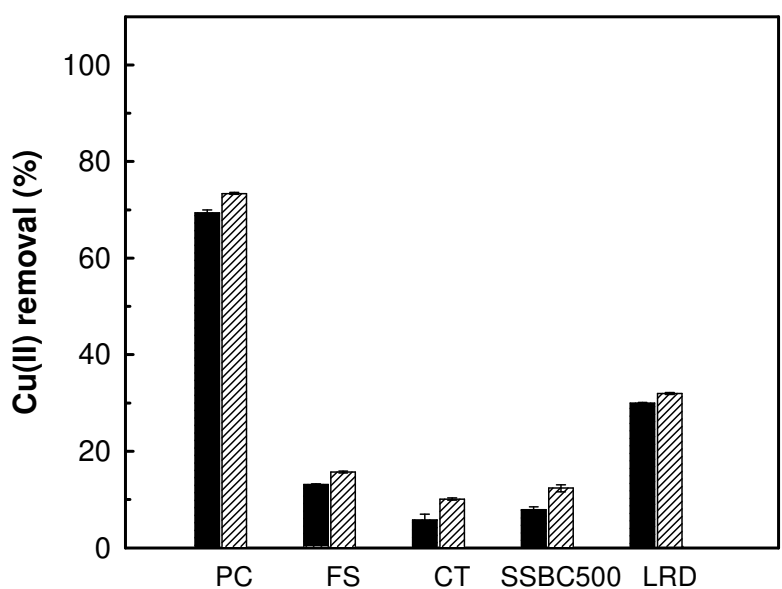

**Fig. 6.** Cu(II) removal efficiencies of PC, FS, CT, SSBC500, and LRD in real (■) and synthetic leachates (▨). Experimental conditions: 6 mM Cu(II), 1.1 mM Zn(II), 1 M $NH_3$, 0.5 M $CO_3^{2-}$, 100 mL $g^{-1}$ L/S ratio, pH 11, RT (20 ± 2 °C), 24 h contact time.

Exhibiting an exceptional adsorption performance, PC has the potential to remediate Cu-$NH_3$ waste streams and subsequently recover Cu. PC is a tannin-rich material containing pyrogallol B-rings with three adjacent –OH groups (Gandini, 1989; Oo et al., 2009). Previous studies have shown that a part of the phenolic –OH groups of tannins could be converted into –$NH_2$ functions by amination (Hashida et al., 2009; Arbenz and Avérous, 2015). In the case of tannins with pyrogallol B-rings, $NH_3$ treatment in the presence of $O_2$ produces 4′-amino-3′,5′-dihydroxybenzene on the B-ring under relatively mild conditions without a catalyst (Hashida et al., 2009). With excess $NH_3$ in the solution, it is probable that amination of the pyrogallol B-rings of PC occurred. The adsorption of Cu-$NH_3$ complexes onto PC likely occurred via a combination of two mechanisms—coordination with the –$NH_2$ functional groups and electrostatic attraction between the positively charged Cu-$NH_3$ complexes and the net negative surface charge of PC at pH > 5.06 ($pH_{PZC}$) caused by the dissociation of O-containing surface groups. X-ray photoelectron spectroscopy (XPS) results (Fig. A.7) show that Cu2p peak (933.1 eV) was detected in the Cu-adsorbed PC aside from the characteristic peaks C1s, N1s, and O1s of raw PC. Additionally, the O1s peak shifted from 533.1 eV to 532.0 eV after Cu(II) adsorption, while the N1s peak shifted from 400.4 eV to 398.7 eV and increased in intensity. These results indicate that N- and O-containing surface groups (Fig. A.7) were responsible for Cu(II) adsorption.

*3.8. Pinecone as a potential Cu(II) biosorbent from Cu-$NH_3$ waste streams*

Pinecones are considered as a forest residue, which further strengthens the need to utilize it as a viable and sustainable biosorbent for Cu(II) removal from leachates and wastewater. The worldwide forest plantation area of *Pinus* species amounts to ~37.4 million ha (Osman, 2013). Assuming that pinecones from 3 to 5 million ha would be harvested annually, with an average yield of 1200 kg/ha (Aniszewska et al., 2018), the annual harvest would amount to 3.6 to 6 million tons of pinecones. The collection of pinecones can present a drawback, but then large-scale collection should be introduced at some point considering the impacts of pinecones on the nutrients and regeneration of forest stands (Aniszewska et al., 2018).

To further enhance the sustainability of this process, subsequent investigations are needed to recover the adsorbed Cu(II) on pinecone. Cu(II) desorption from pinecone has been addressed in some studies, however, desorption needs to be carefully performed to maintain a balance between efficient Cu(II) up-concentration and pinecone condition preservation for the subsequent adsorption-desorption cycles (Ofomaja et al., 2010; Martín-Lara et al., 2016). Although a destructive process, combustion of the Cu(II)-loaded pinecone is also a feasible option to recover Cu. The objective of such approach is to obtain a Cu-rich ash that could be subsequently processed in a metal processing plant (Fomina and Gadd, 2014). For this, it is important to determine the speciation of Cu to ensure it does not form volatile compounds during the decomposition process (Almendros et al., 2015).

4. **Conclusions**

Overall, this study has demonstrated that a top-down approach to biosorbent selection can efficiently highlight an optimal biosorbent for a particular complex waste stream. From the screen of 23 biosorbents under varying pH, $NH_3$, Zn and $CO_3^{2-}$ concentrations, key biosorbents were identified as PC, FS, CT, SSBC500, and LRD. Within the typically observed pH range of ammoniacal waste streams (i.e., 9-12), the highest Cu(II) adsorption was obtained at pH 11, while the minimum was observed at around pH 9, except for PC, which was identified as a high-performing and robust biosorbent with no marked loss in Cu(II) removal between pH 5 and 12. PC was Cu-selective even in the presence of Zn(II) and $CO_3^{2-}$.

PC is a natural and low-cost material, but it has proven its potential in adsorbing Cu(II) for possible Cu(II) recovery from the increasing volumes of complex Cu-$NH_3$ waste streams. Not only it promotes an eco-friendly approach and stability over large pH range relevant to treat $(NH_4)_2CO_3$ bearing effluents, but also provides a new route to recycle Cu for potential use in the metallurgical industries. This study demonstrated the potential of a fully integrated metal recovery technology (i.e., bioleaching followed by biosorption), which could be an effective approach to elemental sustainability. With the increasing environmental awareness of the scientific and engineering community, policymakers, and public in general, the advantages of biosorption are becoming more evident, making it more competitive than the existing technologies for Cu removal and recovery.

**Acknowledgements**

The authors acknowledge Prof. Olivier De Clerck (UGent Phycology Laboratory) for helping with the collection of the macroalgae *Fucus spiralis*; Prof. Frederik Ronsse and Stef Ghysels (UGent Thermochemical Conversion of Biomass research group) for providing the lignin-rich digested stillage; and Prof. Loretta Li (UBC Environmental Engineering Group) for allowing us to use their laboratory for some of the experiments.

**Funding sources**

This work was supported by European Union's Horizon 2020 research and innovation program (METGROW+, grant number 690088) and UGent Special Research Fund (BOF).

**Appendix A. Supplementary Data**

The supplementary data of this work can be found in the online version of the paper. This document contains Texts A.1–A.3, Tables A.1–A.2 and Figs. A.1–A.7.

# Appendix A. Supplementary Data

**Text A.1**

**Data analysis**

The Cu(II) removal efficiency (Eq. A.1) and adsorption capacity (Eq. A.2) of the biosorbents were calculated as follows:

$$R = \frac{C_0 - C_e}{C_0} \times 100 \qquad (A.1)$$

$$q = \frac{(C_0 - C_e) \times V}{m} \qquad (A.2)$$

Where $R$ is the Cu(II) removal efficiency (%), $q$ is the amount of Cu(II) adsorbed per unit mass of the biosorbent (mmol g$^{-1}$), $C_0$ and $C_e$ are the initial and equilibrium concentrations of Cu(II) (mM) respectively, $V$ is the volume of the Cu-NH$_3$ solution (L), and $m$ is the mass of the biosorbent (g).

Langmuir (Eq. A.3) and Freundlich (A.4) isotherm models were used to fit the experimental data of the biosorbents at different initial Cu(II) concentrations.

$$q_e = \frac{q_{max} b C_e}{1 + b C_e} \qquad (A.3)$$

$$q_e = K_f C_e^{1/n} \qquad (A.4)$$

Where $q_e$ is the amount of Cu(II) adsorbed per unit mass of the biosorbent at equilibrium (mmol g$^{-1}$), $q_{max}$ is the maximum adsorption capacity (mmol g$^{-1}$), $b$ is the Langmuir adsorption equilibrium constant (L mmol$^{-1}$), $C_e$ is the equilibrium Cu(II) concentration (mM), $K_f$ is the Freundlich constant (mmol g$^{-1}$)(L mmol$^{-1}$)$^{1/n}$, and 1/n is a dimensionless parameter that varies between 0 and 1.

The selectivity quotients (Eq. A.5) of the biosorbents were calculated as follows:

$$K_{Cu/Zn} = \frac{K_d^{Cu}}{K_d^{Zn}} \qquad (A.5)$$

Where $K_d^{Cu}$ and $K_d^{Zn}$ are the distribution coefficients for Cu and Zn calculated using Eq. A.6.

$$K_d = \frac{100 - X}{X} \cdot \frac{V}{m} \tag{A.6}$$

Where *X* is the equilibrium concentration of either Cu(II) or Zn(II) in solution expressed as a percentage of the initial concentration, *V* is the volume of the solution (mL), and *m* is the mass of the biosorbent (g).

**Text A.2**

**Biosorbent characterizations**

*Fourier transform infrared spectroscopy*

Fourier transform infrared spectroscopy (FTIR) measurements were performed using a Thermo Scientific Nicolet 6700 FT-IR Spectrometer. Approximately 15 mg of dried and ground biosorbent was placed on top of a potassium bromide (KBr) disc for the analysis. FTIR spectra were recorded within the wavenumber range of 400 to 4000 cm$^{-1}$ with spectral resolution of 4 cm$^{-1}$ and 256 scans. The background obtained from the scan of pure KBr was automatically subtracted from each sample spectrum.

*Pore structure characterization*

The pore structure of each biosorbent was characterized by nitrogen adsorption-desorption measurements carried out at 77 K using a Micromeritics Tristar II 3020 apparatus. The samples were degassed at 60 °C under vacuum for at least 30 h prior to measurement. The BET specific surface area ($S_{BET}$) and the total pore volume ($V_T$) of the biosorbents were determined using the BET method.

*Point of zero charge (pH$_{PZC}$) determination*

The pH$_{PZC}$ of the biosorbents were determined using a batch equilibrium method described by Faria et al. (2004). Aliquots of 40 mL of 0.01 M NaCl were transferred to a series of 50-mL polypropylene tubes. The initial pH of each solution was adjusted to a value from 3 to 11 with one pH unit increment. Then, 0.12 g of the biosorbent was added to each tube. The suspensions were shaken at 115 rpm for 48 h in an orbital shaker to allow equilibrium. Afterwards, the final pH of the solutions was measured. For each biosorbent, the initial pH value (pH$_0$) was plotted versus the difference between the initial and final pH values (pH$_0$ – pH$_f$) and the pH$_{PZC}$ was calculated as the point where the curve intersects the x-axis (i.e., pH$_0$ = pH$_f$).

*Acidic and basic surface properties*

The surface groups of the biosorbents giving the acidic and basic surface properties were estimated using Boehm titration (Boehm, 2002). Each biosorbent (0.4 g) was mixed with 40 mL of the following solutions separately: 0.1 M $NaHCO_3$, 0.05 M $Na_2CO_3$, 0.1 M NaOH, and 0.1 M HCl. The suspensions were shaken in an orbital shaker at 115 rpm for 48 h. Subsequently, 10 mL of each suspension was titrated with either 0.05 M HCl or NaOH depending on the starting solution. To calculate the concentrations of the surface oxygen groups, it was assumed that (1) $NaHCO_3$ reacts with carboxylic groups only, (2) $Na_2CO_3$ reacts with both carboxylic and lactonic groups, and (3) NaOH reacts with carboxylic, lactonic, and phenolic groups. The quantification of basic sites was also calculated from the amount of HCl that reacted with the biosorbents.

*X-ray photoelectron spectroscopy (XPS)*

XPS analyses of raw and Cu-adsorbed PC were performed on S-Probe Monochromatized XPS spectrometer (Surface Science Instruments, VG) with monochromatic AlKα X-ray (1486.6 eV, 200 W) source. The low-resolution spectra were collected with a pass energy $E_p = 140.9$ eV and energy steps $E_s = 0.24$ eV, while the high-resolution spectra with $E_p = 140.9$ eV and $E_s = 0.15$ eV. The XPS spectra were analyzed using CasaXPS Software.

**Text A.3**

**Chemical speciation modeling**

The speciation of Cu(II) and Zn(II) in the presence of $NH_3$ and $CO_3^{2-}$ in aqueous solutions was determined using the Hydra-Medusa software. Hydra contains a database with the logarithm of equilibrium constants (log K) at 25 °C, while Medusa creates the different equilibrium diagrams (e.g., species fraction, logarithmic, and predominance area diagrams) (Puigdomenech, 2013). The initial concentration of each component in the defined system was used as the input to plot the fraction of the predominant metal species as a function of the solution pH.

**Table A.1.** Main properties of the biosorbents pinecone (PC), *Fucus spiralis* (FS), chitosan (CT), sewage sludge biochar (SSBC500), and lignin-rich digested stillage (LRD).

|  | PC | FS | CT | SSBC500 | LRD |
|---|---|---|---|---|---|
| $pH_{PZC}$ | 5.06 | 5.97 | 6.60[†] | 7.49 | 6.34 |
| $S_{BET}$ (m$^2$ g$^{-1}$) | 0.6 | ~0 | 0.8 | 3.2 | 3.8 |
| $V_T$ (cm$^3$ g$^{-1}$) | 0.001 | < 0.001 | 0.002 | 0.007 | 0.014 |
| **Surface sites** | | | | | |
| Carboxylic | 0.38 | 0.12 | N.D.[‡] | 0.11 | 0.96 |
| Lactonic | 0.23 | 0.39 | N.D. | 0.03 | 0.15 |
| Phenolic | 0.67 | 0.36 | N.D. | 0.79 | 0.01 |
| Basic sites | 0.16 | 1.38 | N.D. | 0.48 | 1.27 |

[†] Average of values reported in literature (Fras et al., 2012; Marques Neto et al., 2013; Caje et al., 2017; Gür et al., 2017)

[‡] N.D. = Not determined

**Table A.2.** Analysis of the FTIR spectra peaks of the raw biosorbents.

**Pinecone (PC)**

| Peak position (cm$^{-1}$) | Assignment |
|---|---|
| 3450 | unbounded –OH group |
| 2933 | aliphatic C–H group stretching |
| 1722 | C=O stretch (non-cyclic esters – lactones) |
| 1612 | C=C (aromatic skeletal mode of lignin) and C=O (carboxylic acids) |
| 1510 | aromatic ring vibrations likely from the lignin fraction of the plant material |
| 1440 | –COO$^-$ (carboxylate salts) |
| 1272 | C–N stretching with amine |
| 1109 | C–O–C functionalities |

*Fucus spiralis* **(FS)**

| Peak position (cm$^{-1}$) | Assignment |
|---|---|
| 3332 | stretching vibration of O–H group |
| 2924 | asymmetric stretch of aliphatic chains (–CH) |
| 1741 | C=O stretch of COOH |
| 1664 | asymmetric C=O |
| 1531 | asymmetric stretching vibrations of COO$^-$ group |
| 1444 | symmetric stretching vibrations of COO$^-$ group |
| 1261 | stretching vibration of the bond C=S which is characteristic of sulfonic acid |
| 1101 | C–O (ether) |
| 820 | S=O stretch |

**Chitosan (CT)**

| Peak position (cm$^{-1}$) | Assignment |
|---|---|
| 3433 | N–H and O–H stretching |
| 2918, 2864 | C–H symmetric stretching |
| 1662 | C=O stretching of amide I (presence of residual *N*-acetyl groups) |
| 1554 | N–H bending of amide II |
| 1385 | CH$_3$ symmetrical deformations |
| 1144 | asymmetric stretching of the C–O–C bridge |
| 1070, 1043 | C–O stretching |

**Sewage sludge biochar (SSBC500)**

| Peak position (cm$^{-1}$) | Assignment |
|---|---|
| 3352 | generally attributed to O–H and secondarily to N–H stretching of various functional groups in H bonding |
| 3049 | aliphatic C–H stretching |
| 1581 | aromatic C=O stretching vibration and the C=C stretching vibration of carboxylic acids esters, ketones, and anhydrides |
| 1150 | organic C–OH stretching of carbohydrates |
| 1047 | C–O stretching vibration |

**Lignin-rich digested stillage (LRD)**

| Peak position (cm$^{-1}$) | Assignment |
|---|---|
| 3360 | aliphatic and phenolic OH-groups |
| 2927 | alkane C–H stretching |
| 1666, 1591, 1506, 1454, 1423 | C=C stretching in the aromatic rings |
| 1317 | C–O–C symmetric stretching |
| 1263, 1228 | C=O stretching |
| 1115, 1063 | C–O–C symmetric stretching |

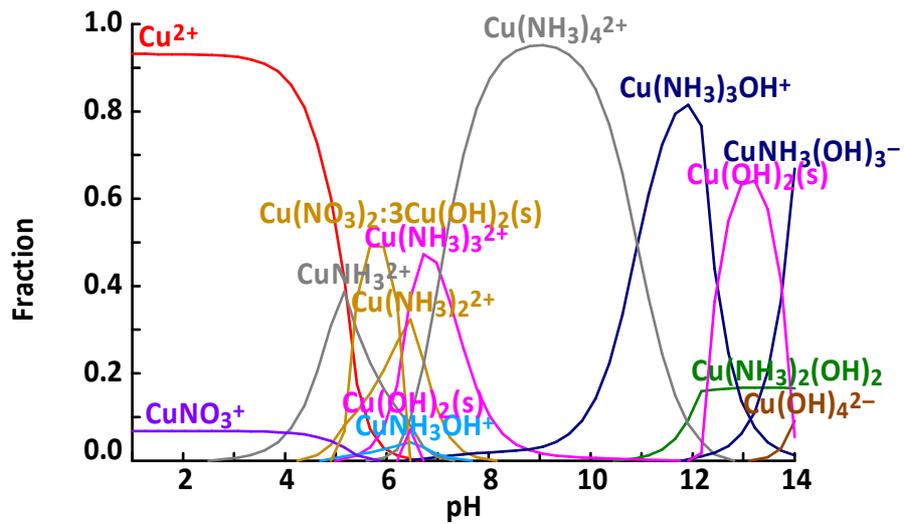

**Fig. A.1.** Chemical speciation of 12 mM Cu(II) in the presence of 1 M NH$_3$ estimated using the Hydra-Medusa software.

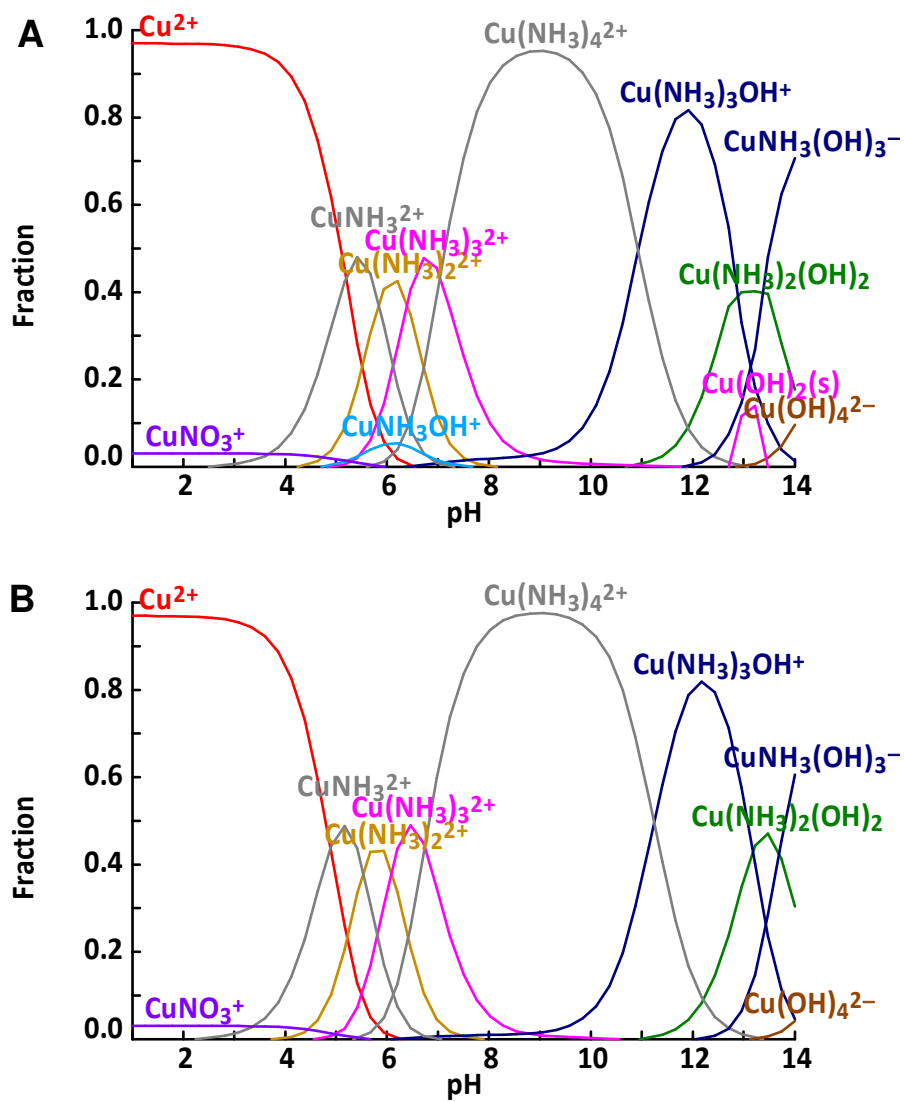

**Fig. A.2.** Chemical speciation of 5 mM Cu(II) in aqueous solution in the presence of (A) 1 M or (B) 2 M NH$_3$ estimated using the Hydra-Medusa software.

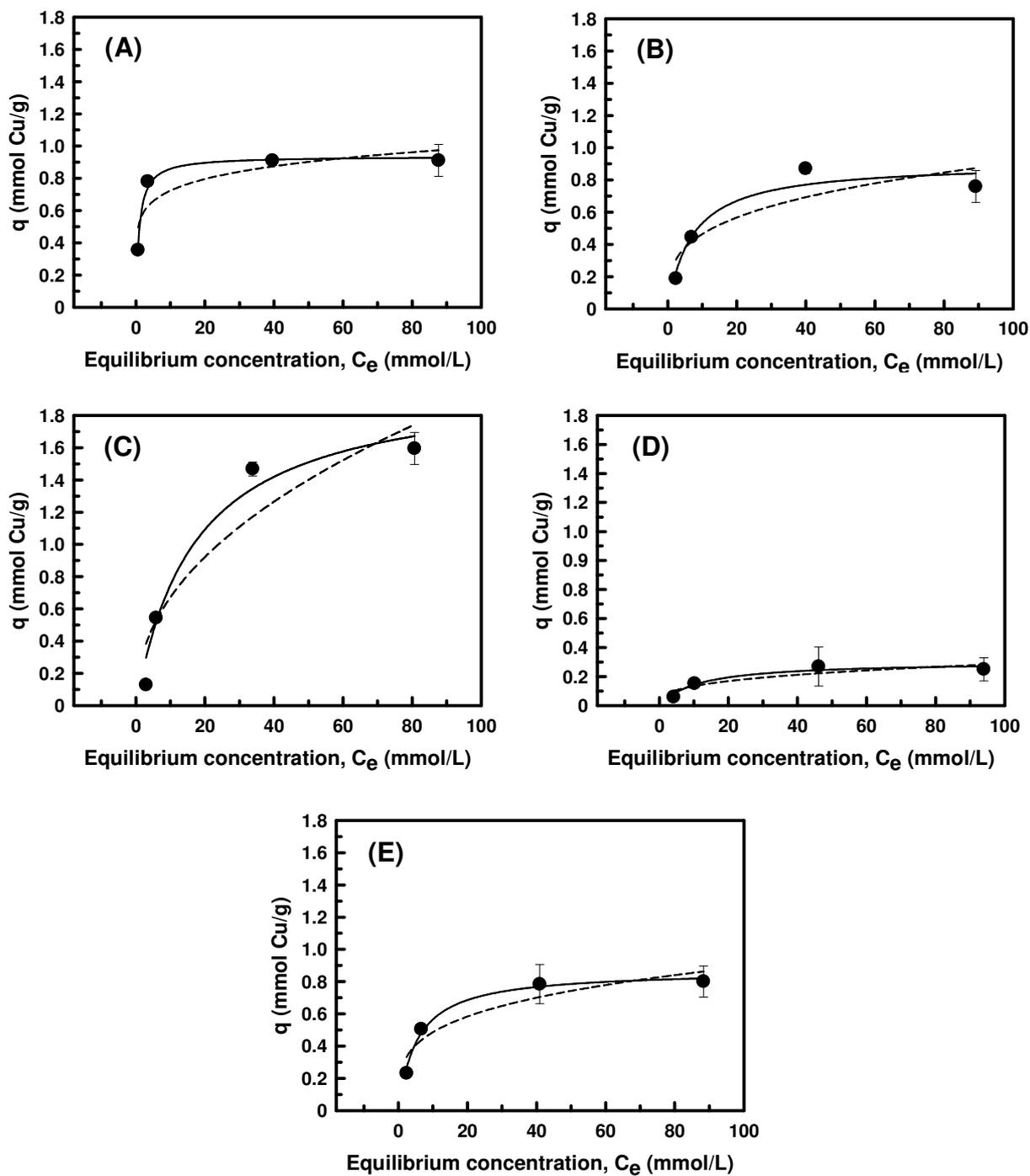

**Fig. A.3.** Langmuir (solid) and Freundlich (dashed) isotherm models of Cu(II) adsorption onto (A) PC, (B) FS, (C) CT, (D) SSBC500, and (E) LRD in the presence of $NH_3$. Experimental conditions: 2 M $NH_3$, 100 mL $g^{-1}$ L/S ratio, pH 11, room temperature (RT), 24 h contact time.

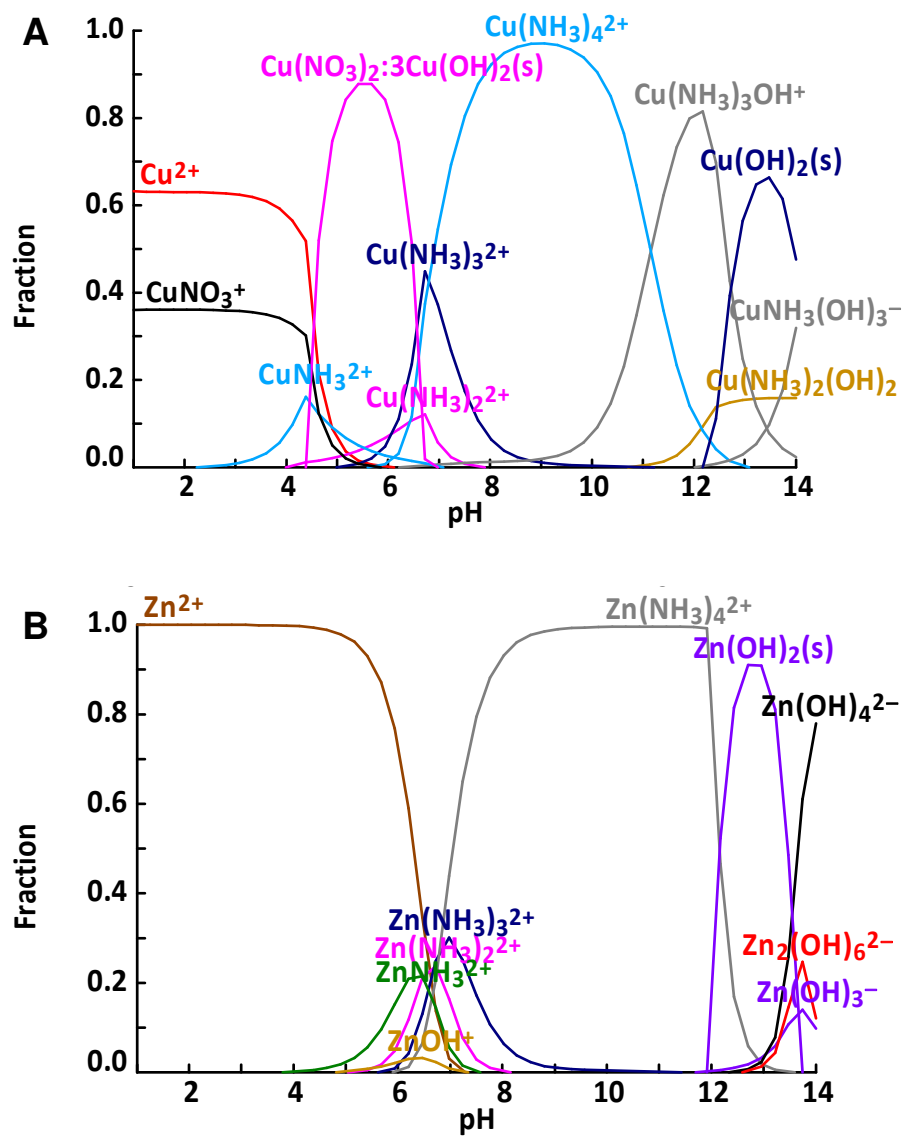

**Fig. A.4.** Chemical speciation of (A) 50 mM Cu(II) in the presence of 50 mM Zn(II) and 2 M NH$_3$ and (B) 50 mM Zn(II) in the presence of 50 mM Cu(II) and 2 M NH$_3$ estimated using the Hydra-Medusa software.

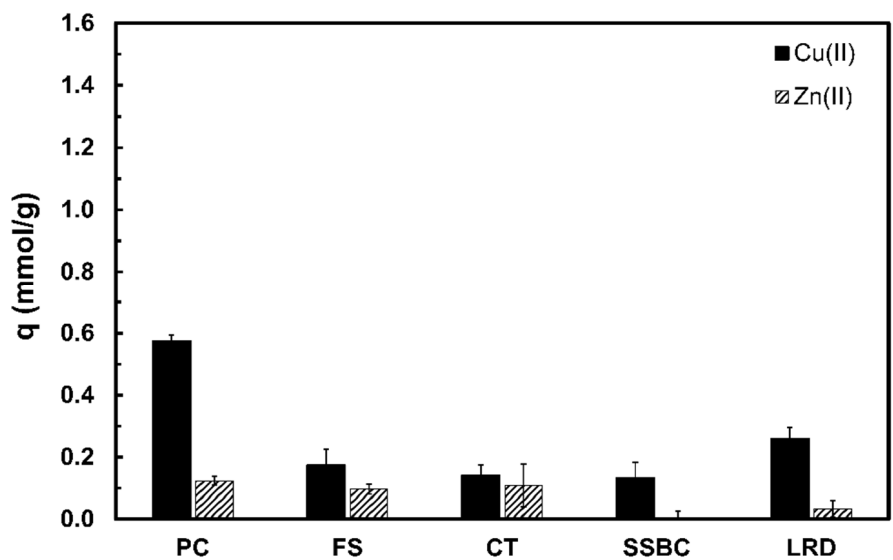

**Fig. A.5.** Cu(II) and Zn(II) adsorption capacities of PC, FS, CT, SSBC500, and LRD in binary metal solutions containing $NH_3$. Experimental conditions: 50 mM Cu(II), 50 mM Zn(II), 2 M $NH_3$, 100 mL g$^{-1}$ L/S ratio, pH 9, RT, 24 h contact time.

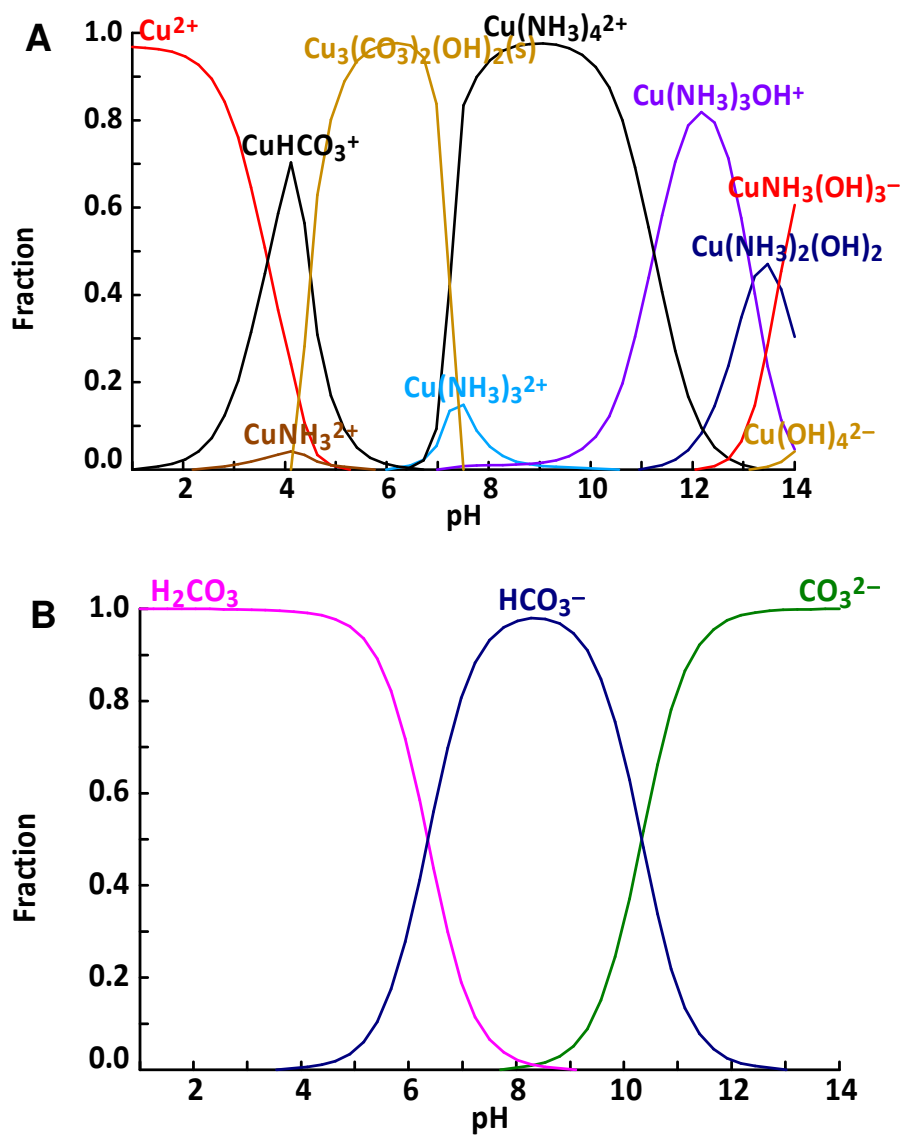

**Fig. A.6.** Chemical speciation of (A) 5 mM Cu(II) in the presence of 2 M $NH_3$ and 1 M $CO_3^{2-}$ and (B) 1 M $CO_3^{2-}$ in the presence of 5 mM Cu(II) and 2 M $NH_3$ estimated using the Hydra-Medusa software.

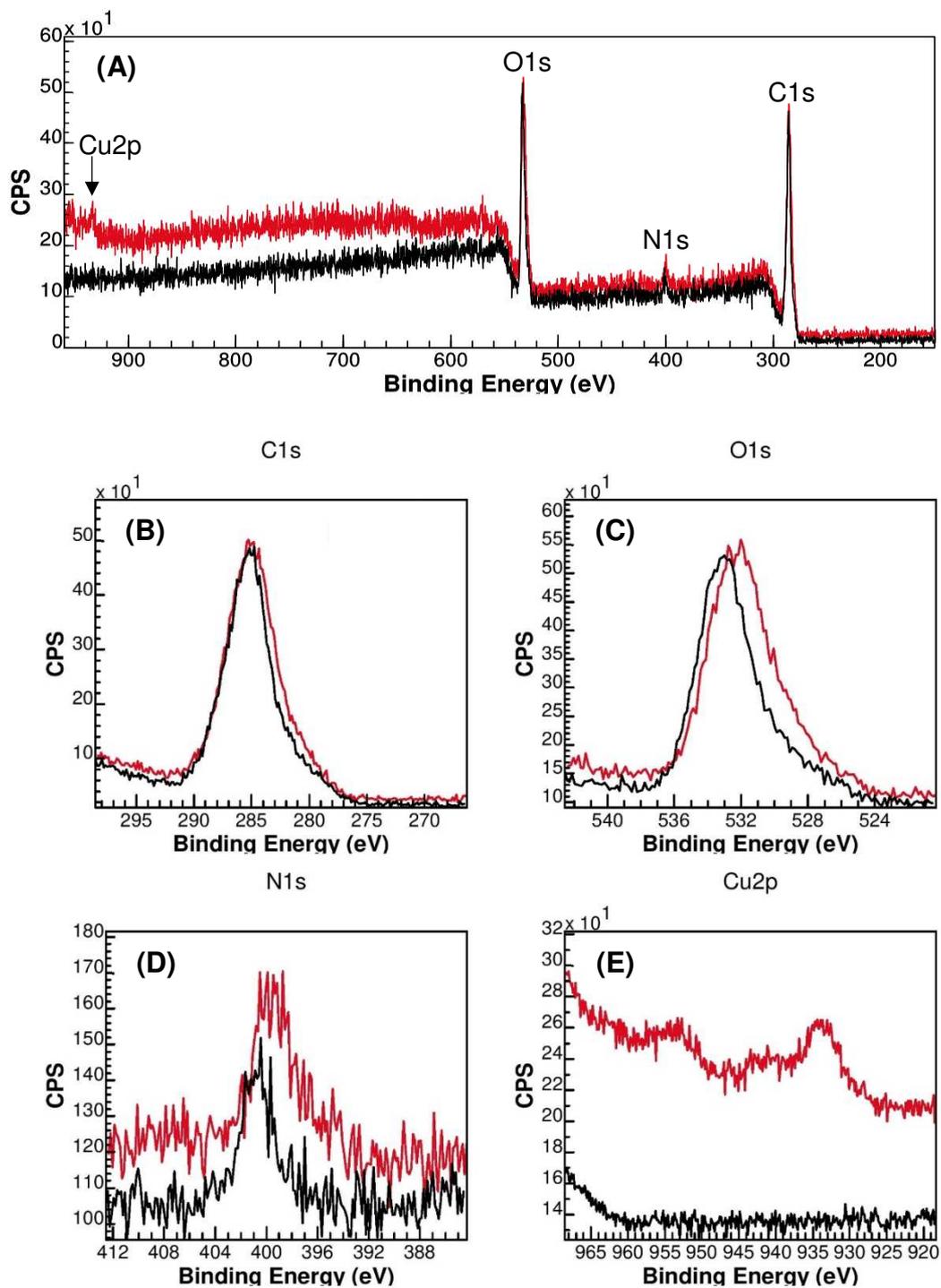

**Fig. A.7.** (A) XPS survey spectra of raw PC (black) and Cu-loaded PC (red) and high-resolution XPS scan spectra over (B) C 1s; (C) O 1s; (D) N 1s; and (E) Cu 2p of raw PC (black) and Cu-loaded PC (red). The spectra are plotted as counts per second (CPS) versus the binding energy. Note: The spectra noise is due to the short acquisition time that had to be used to avoid the X-ray photoreduction of Cu(II) ions.